\documentclass[conference]{IEEEtran}
\IEEEoverridecommandlockouts
\usepackage{cite}
\usepackage{amsmath,amssymb,amsfonts}
\usepackage{algorithmic}
\usepackage{graphicx}
\usepackage{textcomp}
\usepackage{xcolor}
\usepackage{url}
\usepackage{multirow}
\def\BibTeX{{\rm B\kern-.05em{\sc i\kern-.025em b}\kern-.08em
    T\kern-.1667em\lower.7ex\hbox{E}\kern-.125emX}}
\begin{document}

\title{Multimodal Real-Time Anomaly Detection and Industrial Applications}

\author{\IEEEauthorblockN{ Aman Verma}
\IEEEauthorblockA{\textit{Department of Electrical Engineering} \\
\textit{IIT Bombay}\\
Mumbai, India \\
22b3929@iitb.ac.in}
\and
\IEEEauthorblockN{ Keshav Samdani}
\IEEEauthorblockA{\textit{Department of Electrical Engineering} \\
\textit{IIT Bombay}\\
Mumbai, India \\
22b3952@iitb.ac.in}
\and
\IEEEauthorblockN{ Mohd. Samiuddin Shafi}
\IEEEauthorblockA{\textit{Department of Electrical Engineering} \\
\textit{IIT Bombay}\\
Mumbai, India \\
22b3912@iitb.ac.in}
}

\maketitle

\begin{abstract}
This paper presents the design, implementation, and evolution of a comprehensive multimodal room monitoring system that integrates synchronized video and audio processing for real-time activity recognition and anomaly detection. We describe two iterations of the system: an initial lightweight implementation using YOLOv8, ByteTrack, and Audio Spectrogram Transformer (AST), and an advanced version incorporating multi-model audio ensembles, hybrid object detection, bidirectional cross-modal attention, and multi-method anomaly detection. The evolution demonstrates significant improvements in accuracy, robustness, and industrial applicability. The advanced system combines three audio models (AST, Wav2Vec2, HuBERT) for comprehensive audio understanding, dual object detectors (YOLO and DETR) for improved accuracy, and sophisticated fusion mechanisms for enhanced cross-modal learning. Experimental evaluation shows the system's effectiveness in both general monitoring scenarios and specialized industrial safety applications, achieving real-time performance on standard hardware while maintaining high accuracy.
\end{abstract}


\section{Introduction}

The increasing demand for intelligent monitoring systems in smart homes, security applications, healthcare facilities, and industrial environments has driven significant research into multimodal perception systems. Traditional single-modality approaches, whether relying solely on visual or audio information, often fail to capture the complete context of activities and events. Visual systems may miss audio cues that provide crucial context, while audio-only systems lack spatial understanding. Multimodal fusion addresses these limitations by combining complementary information from different sensory modalities, enabling more robust and comprehensive scene understanding.

This paper presents a comprehensive journey through the development of a multimodal room monitoring system, documenting both an initial lightweight implementation and its evolution into an advanced industrial-grade solution. The initial system, designed for proof-of-concept and educational purposes, demonstrates the feasibility of real-time multimodal fusion using state-of-the-art but computationally efficient models. The advanced system extends these capabilities with sophisticated ensemble methods, hybrid detection strategies, and specialized industrial monitoring features.

The initial system integrates YOLOv8 for object detection, ByteTrack for multi-object tracking, and Audio Spectrogram Transformer (AST) for audio scene classification, all connected through a lightweight cross-modal transformer architecture. This foundation proved effective for basic activity recognition tasks, demonstrating the potential of synchronized multimodal processing. However, practical deployment revealed limitations in accuracy, robustness, and specialized application support.

The advanced system addresses these limitations through several key innovations. First, a multi-model audio ensemble combines AST, Wav2Vec2, and HuBERT models, each specializing in different aspects of audio understanding—general classification, speech recognition, and acoustic scene analysis. Second, a hybrid object detection system pairs the speed of YOLO with the accuracy of DETR, using cross-detector non-maximum suppression to eliminate duplicates. Third, bidirectional cross-modal attention allows visual and audio features to mutually inform each other, rather than simple concatenation. Finally, multi-method anomaly detection combines statistical, learned, and event-based approaches for comprehensive anomaly identification.

The main contributions of this work are: (1) a detailed architectural evolution from basic to advanced multimodal fusion, (2) a multi-model audio ensemble approach for robust audio understanding, (3) a hybrid object detection system balancing speed and accuracy, (4) bidirectional cross-modal attention mechanisms for enhanced fusion, (5) multi-method anomaly detection suitable for industrial applications, and (6) comprehensive experimental evaluation demonstrating real-time performance and practical applicability.

\section{Related Work}

Multimodal fusion for activity recognition and scene understanding has been extensively studied in recent literature. Early approaches primarily employed late fusion strategies, where separate models process each modality independently and their predictions are combined at the final stage \cite{ref1}. While simple to implement, these methods fail to capture cross-modal interactions and temporal dependencies that are crucial for understanding complex activities.

The advent of transformer architectures revolutionized multimodal learning. The Audio-Visual Transformer (AVT) \cite{ref2} demonstrated the effectiveness of self-attention mechanisms for learning cross-modal representations. Similarly, multimodal transformers have shown remarkable success in video understanding tasks \cite{ref3}, where temporal and cross-modal dependencies are naturally modeled through attention mechanisms.

Object detection has seen continuous improvements with YOLO variants \cite{ref4} achieving real-time performance while maintaining accuracy. ByteTrack \cite{ref5} introduced effective association strategies for multi-object tracking, handling occlusions and re-identification through both high and low confidence detections. DETR \cite{ref6} demonstrated that transformer-based detection can achieve state-of-the-art accuracy, though at higher computational cost.

Audio classification has been transformed by Audio Spectrogram Transformers \cite{ref7}, which adapt vision transformer architectures for audio processing. Wav2Vec2 \cite{ref8} introduced self-supervised learning for speech recognition, while HuBERT \cite{ref9} improved upon this with better acoustic representations. However, most systems employ single models rather than ensembles.

Anomaly detection in multimodal settings has been approached through various methods. Statistical approaches using z-scores and thresholds are simple but limited \cite{ref10}. Autoencoder-based methods learn normal patterns and detect deviations \cite{ref11}. Event-based methods leverage classification outputs to identify anomalous events \cite{ref12}. Our work combines all three approaches for comprehensive anomaly detection.

Industrial monitoring systems have specific requirements including real-time performance, high reliability, and specialized event detection. Previous work has focused on single-modality industrial monitoring \cite{ref13}, but multimodal approaches remain underexplored. Our system addresses this gap with specialized industrial monitoring modes and safety-focused event detection.

\section{System Architecture Evolution}

\subsection{Initial System Architecture}

\begin{figure}[!t]
\centering
\includegraphics[width=\columnwidth]{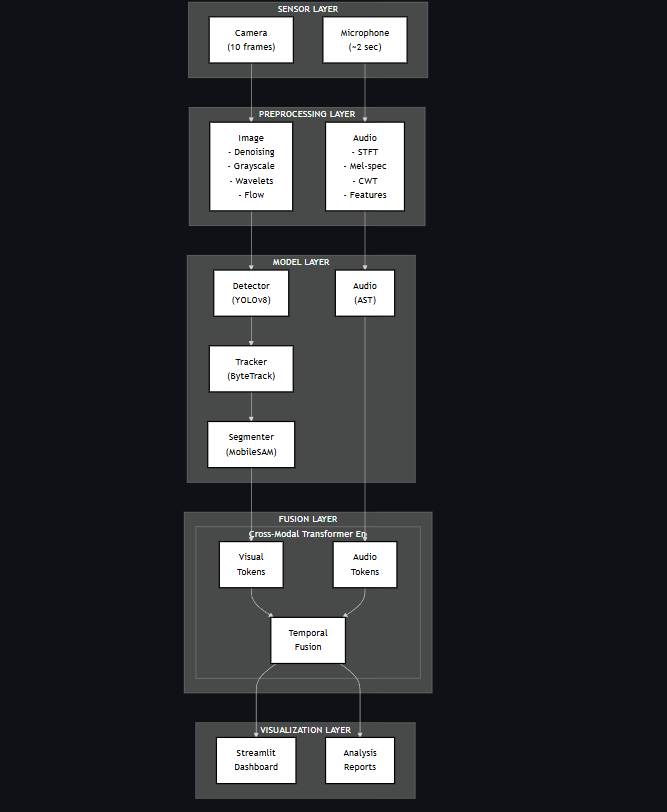}
\caption{Initial system architecture showing the four main layers: sensor, preprocessing, model, and fusion layers. This lightweight design prioritizes real-time performance and computational efficiency.}
\label{fig:architecture_basic}
\end{figure}

The initial system was designed with simplicity and real-time performance as primary goals. The architecture consists of four main layers: sensor layer, preprocessing layer, model layer, and fusion layer. Figure~\ref{fig:architecture_basic} illustrates the overall system architecture of the initial implementation.

The sensor layer captures synchronized multimodal data from standard camera and microphone devices. The camera captures a burst of 10 frames at approximately 20 frames per second, providing sufficient temporal resolution for motion analysis while maintaining manageable computational load. The microphone records approximately 2 seconds of audio at 16 kHz sampling rate, balancing frequency resolution with processing requirements. Crucially, precise timestamps are recorded for each frame, enabling accurate temporal alignment during fusion—a requirement that becomes increasingly important as system complexity grows.

The preprocessing layer performs essential signal conditioning for both modalities. Visual preprocessing begins with fast non-local means denoising, which reduces noise while preserving edges and fine details critical for object detection. Frames are then converted to grayscale for optical flow computation and wavelet analysis, reducing computational complexity while maintaining essential information. Multi-resolution texture analysis using Daubechies-2 wavelets with 2-level decomposition extracts energy features that capture texture characteristics at different scales. Finally, TV-L1 optical flow algorithm computes dense motion vectors between consecutive frames, enabling motion analysis that complements object detection.

Audio preprocessing extracts multiple complementary representations. Short-Time Fourier Transform (STFT) provides standard time-frequency representation with configurable window size and hop length. Mel-scale spectrogram captures perceptually relevant frequency information using 64 mel bands, aligning with human auditory perception. Continuous Wavelet Transform (CWT) using Morlet wavelets provides time-scale representation particularly effective for non-stationary signal analysis. Statistical features including zero-crossing rate (ZCR), spectral centroid, bandwidth, and rolloff are extracted to provide compact yet informative representations.

The model layer employs carefully selected state-of-the-art models optimized for real-time performance. YOLOv8 nano variant performs object detection with 80 COCO classes, configured with confidence threshold of 0.3 and IoU threshold of 0.5 for non-maximum suppression. This configuration balances detection sensitivity with false positive reduction. ByteTrack algorithm maintains persistent track IDs across frames by associating detections using both high and low confidence predictions, enabling robust tracking through occlusions and temporary detection failures—a critical capability for real-world deployment.

Instance segmentation is performed using MobileSAM or FastSAM models, providing pixel-level object boundaries that enhance understanding beyond bounding boxes. The Audio Spectrogram Transformer (AST) from HuggingFace, pre-trained on AudioSet, processes mel-spectrogram patches using a vision transformer architecture adapted for audio. The model outputs probabilities over 527 AudioSet classes, enabling comprehensive audio event classification.

The fusion layer employs a lightweight cross-modal transformer architecture designed for efficiency. Visual tokens are constructed from per-frame features including bounding box count, mean detection confidence, and wavelet energy. Audio tokens incorporate spectral features including zero-crossing rate, spectral centroid, bandwidth, and rolloff. These tokens are projected to a common hidden dimension of 128, combined via residual addition, and processed through a 2-layer transformer encoder with 4 attention heads. Temporal pooling followed by a classification head outputs binary activity recognition (static vs. moving).

\subsection{Advanced System Architecture}

\begin{figure}[!t]
\centering
\includegraphics[width=\columnwidth]{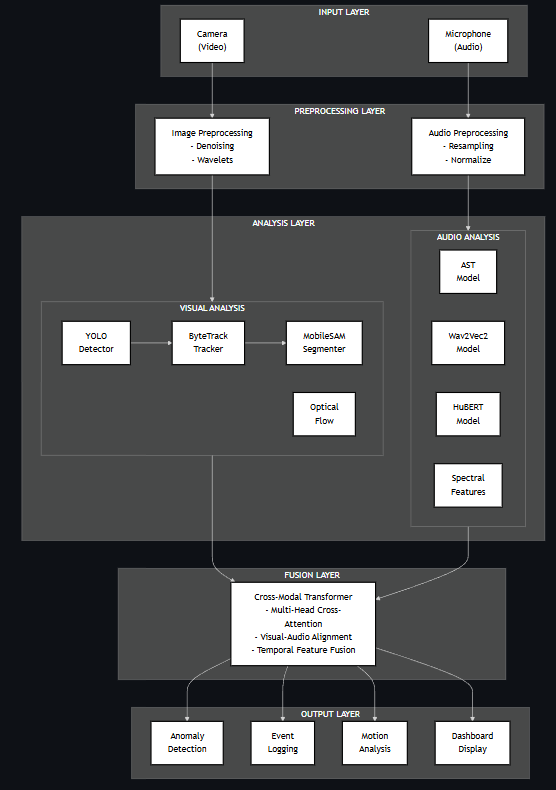}
\caption{Advanced system architecture incorporating multi-model audio ensemble, hybrid object detection, bidirectional cross-modal attention, and multi-method anomaly detection. The system maintains real-time performance while significantly improving accuracy and robustness.}
\label{fig:architecture_advanced}
\end{figure}

The advanced system extends the initial architecture with sophisticated enhancements addressing accuracy, robustness, and industrial applicability. Figure~\ref{fig:architecture_advanced} illustrates the enhanced architecture with new components and improved data flow.

The most significant enhancement is the multi-model audio ensemble, which combines three specialized audio models. The Audio Spectrogram Transformer (AST) provides general audio event classification, processing mel-spectrogram patches through a vision transformer adapted for audio. Wav2Vec2 specializes in speech recognition and transcription, employing self-supervised learning to extract robust acoustic representations. HuBERT focuses on acoustic scene understanding, providing complementary features that enhance overall audio comprehension. A fusion layer combines embeddings from all three models (each 768-dimensional) into a unified 256-dimensional representation, learning optimal combinations through training.

The hybrid object detection system addresses the speed-accuracy tradeoff by combining YOLO and DETR. YOLO provides fast, efficient detection suitable for real-time processing, while DETR offers superior accuracy for complex scenes through transformer-based detection. Cross-detector non-maximum suppression eliminates duplicate detections, ensuring that the system benefits from both models' strengths without redundancy. This hybrid approach is particularly effective in industrial environments where both speed and accuracy are critical.

The fusion architecture is significantly enhanced with bidirectional cross-modal attention. Rather than simple concatenation or residual addition, the advanced system employs multi-head cross-attention mechanisms where visual features attend to audio features and vice versa. This bidirectional attention allows each modality to inform the other—visual context can help disambiguate audio events (e.g., seeing a crash helps identify crash sounds), while audio cues can guide visual attention (e.g., sounds help locate visual events). The architecture includes 4 transformer layers, each with 8 attention heads and 256 hidden dimensions, providing sufficient capacity for complex cross-modal learning.

Anomaly detection is implemented through multiple complementary methods. Statistical anomaly detection uses z-scores based on pixel intensity statistics, providing fast detection of unusual visual patterns. Autoencoder-based detection employs a convolutional autoencoder trained on normal data, with high reconstruction error indicating anomalies—this learned approach captures complex patterns that statistical methods might miss. Audio anomaly detection combines energy and spectral centroid analysis, detecting unusual audio characteristics. Event-based detection leverages classification outputs, identifying known anomaly classes from both visual and audio models. These methods are combined to provide comprehensive anomaly identification suitable for industrial safety applications.

\subsection{Industrial Monitoring Enhancements}

The advanced system includes specialized industrial monitoring capabilities addressing safety-critical applications. Industrial mode focuses on specific event categories including machine operation, tool usage, smoke detection, fire detection, leak detection, and structural damage. The system employs specialized audio classification tuned for industrial sounds, enabling detection of machinery malfunctions, safety equipment activation, and hazardous conditions.

Real-time background processing ensures responsive user interface while maintaining comprehensive analysis. A separate thread handles frame analysis, preventing UI freezing during heavy computation. Queue-based architecture enables smooth processing with configurable buffer sizes, maintaining real-time video display while performing detailed analysis in the background. This architecture is critical for industrial applications where both real-time monitoring and detailed analysis are required.

\section{Detailed System Components}

\subsection{Multi-Model Audio Analysis}

The advanced audio analysis system represents a significant departure from single-model approaches. The Audio Spectrogram Transformer processes mel-spectrogram patches through a vision transformer architecture, providing general audio event classification across 527 AudioSet classes. This model excels at identifying environmental sounds, music, and general audio events, providing broad coverage of audio scenes.

Wav2Vec2 employs self-supervised learning to extract robust acoustic representations, specializing in speech recognition and transcription. The model processes raw audio waveforms, learning representations that capture phonetic and linguistic information. This capability is particularly valuable in industrial settings where speech commands, alarms, or verbal warnings may be present.

HuBERT focuses on acoustic scene understanding, providing features that capture environmental context and acoustic characteristics. The model's representations complement AST's event classification and Wav2Vec2's speech focus, providing a third perspective on audio understanding.

The fusion of these three models is achieved through a learned combination layer. Each model produces 768-dimensional embeddings, which are concatenated and processed through a linear projection layer that reduces dimensionality to 256 while learning optimal combinations. This fusion approach allows the system to leverage each model's strengths while maintaining computational efficiency. The fused embeddings capture comprehensive audio understanding, enabling more accurate cross-modal fusion and better anomaly detection.

\subsection{Hybrid Object Detection}

The hybrid detection system addresses fundamental limitations of single-detector approaches. YOLO provides exceptional speed, processing frames in 100-200ms on CPU, making it ideal for real-time applications. However, YOLO can struggle with complex scenes, small objects, and unusual viewpoints. DETR, while slower (typically 500-1000ms per frame), provides superior accuracy through transformer-based detection that naturally handles complex spatial relationships.

The system employs both detectors in parallel, with YOLO providing primary detections and DETR providing supplementary high-confidence detections. Cross-detector non-maximum suppression eliminates duplicates by computing IoU between detections from different models and removing overlapping boxes. This approach ensures that the system benefits from YOLO's speed while incorporating DETR's accuracy where needed.

The hybrid approach is particularly effective in industrial environments where both speed and accuracy are critical. Real-time monitoring requires fast processing, but safety-critical applications demand high accuracy. The hybrid system provides a practical solution, using YOLO for most frames and DETR for complex scenes or when high confidence is required.

\subsection{Bidirectional Cross-Modal Fusion}

The advanced fusion architecture employs bidirectional cross-modal attention, a significant enhancement over simple feature concatenation. The architecture processes visual and audio tokens through separate projection layers, mapping them to a common 256-dimensional space. Positional embeddings are added to preserve temporal information, critical for understanding activities that unfold over time.

The core innovation is the multi-head cross-attention mechanism. In each transformer layer, visual features attend to audio features, allowing visual context to inform audio understanding. Simultaneously, audio features attend to visual features, enabling audio cues to guide visual attention. This bidirectional attention creates a rich cross-modal representation where each modality enhances the other's understanding.

The attention mechanism is implemented as:
\begin{equation}
\text{Attention}(Q, K, V) = \text{softmax}\left(\frac{QK^T}{\sqrt{d_k}}\right)V
\end{equation}
where for visual-to-audio attention, $Q$ comes from visual tokens, $K$ and $V$ from audio tokens, and vice versa for audio-to-visual attention.

The architecture includes 4 transformer layers, each with 8 attention heads and feed-forward networks with 1024 hidden dimensions. Layer normalization and residual connections ensure stable training and effective gradient flow. After processing through all layers, sequence pooling (mean over temporal dimension) aggregates information, followed by task-specific heads for motion classification and event classification across 32 categories.

\subsection{Multi-Method Anomaly Detection}

Anomaly detection employs three complementary methods, each addressing different aspects of anomaly identification. Statistical anomaly detection computes z-scores based on pixel intensity distributions, detecting unusual visual patterns through simple yet effective statistical analysis. This method is fast and requires no training, making it suitable for real-time applications.

Autoencoder-based detection employs a convolutional autoencoder trained on normal data. The encoder compresses input frames to a latent representation, while the decoder reconstructs the original frame. High reconstruction error indicates anomalies—patterns not well-represented in the training data. This learned approach captures complex patterns that statistical methods might miss, but requires training data and computational resources.

Audio anomaly detection combines energy analysis and spectral centroid tracking. Unusual energy levels or frequency characteristics indicate audio anomalies. This method is particularly effective for detecting sudden loud sounds, unusual frequency content, or silence where sound is expected.

Event-based detection leverages classification outputs from both visual and audio models. Known anomaly classes (e.g., fire, smoke, damage) are identified through classification, providing semantic understanding of anomalies. This method complements statistical and learned approaches by providing interpretable anomaly identification.

The system combines these methods through weighted scoring, where each method contributes to a final anomaly score. Thresholds are configurable, allowing adaptation to different environments and requirements. Detected anomalies trigger alerts and save artifacts (frames and audio snippets) for later analysis.

\section{Implementation Details}

\subsection{Data Capture and Synchronization}

Precise temporal synchronization is fundamental to effective multimodal fusion. The system records timestamps for each video frame at capture time, ensuring accurate temporal alignment. Audio samples are aligned to frames using windowed extraction, where each frame timestamp corresponds to an audio window centered at that timestamp. The window size is calculated as the total audio duration divided by the number of frames, ensuring consistent temporal coverage.

The synchronization algorithm handles edge cases including frame sequences shorter than audio duration, variable frame rates, and audio buffer boundaries. Padding is applied when necessary to maintain consistent window sizes, critical for batch processing and model inference. This precise alignment ensures that visual and audio features correspond to the same temporal context, enabling accurate cross-modal fusion.

\subsection{Real-Time Processing Architecture}

The advanced system employs a sophisticated real-time processing architecture balancing responsiveness and comprehensive analysis. The main thread handles user interface updates and real-time video display, ensuring responsive interaction. A separate background thread performs detailed frame analysis, including object detection, tracking, segmentation, and feature extraction.

Queue-based architecture enables smooth processing with configurable buffer sizes. Frames are added to a processing queue as they are captured, while the background thread continuously processes queued frames. This architecture prevents UI freezing during heavy computation while maintaining real-time video display. The queue size is dynamically adjusted based on processing speed, preventing buffer overflow while minimizing latency.

Audio processing employs callback-based streaming, where audio samples are continuously captured and buffered. The callback function adds audio blocks to a buffer, which is periodically processed to extract features and perform classification. This streaming approach ensures continuous audio analysis without blocking the main thread.

Model caching is implemented using Streamlit's resource caching mechanism, ensuring that models are loaded once and reused across multiple inference calls. This significantly reduces initialization time and memory usage, critical for real-time applications. GPU memory management includes automatic cleanup and configurable batch sizes to prevent out-of-memory errors.

\subsection{User Interface and Visualization}

The Streamlit-based user interface provides comprehensive visualization and interactive control. Real-time video feed displays annotated frames with bounding boxes, track IDs, optical flow arrows, and segmentation masks. Color coding distinguishes different object classes and track persistence, enhancing visual understanding.

Audio analysis is visualized through mel-spectrograms, CWT scalograms, and classification results. Top-k predictions are displayed with confidence scores, enabling users to understand system confidence and potential ambiguities. Event logs provide timestamped records of detected events, with filtering and sorting capabilities for analysis.

Anomaly trends are visualized through time-series charts showing anomaly scores over time. This visualization helps identify patterns and trends in anomaly detection, valuable for understanding system behavior and tuning thresholds. Statistics dashboard provides summary metrics including detection counts, average confidence, and processing performance.

Industrial monitoring mode includes specialized visualizations focusing on safety-critical events. Alert system provides immediate notification of detected anomalies, with configurable severity levels and notification methods. Saved anomaly artifacts (frames and audio snippets) are organized by timestamp and anomaly type, enabling detailed post-analysis.

\section{Experimental Evaluation}

\subsection{Experimental Setup}

We tried multiple situations like collisions, spraying, fire and certain audio visual situations where the multi-modality of the paper is utilized

\subsection{Object Detection and Tracking Performance}

This is the case of finger snapping which was done with a basic room setting case we have also included metallic collision also 

Figure 3 shows how the flow vectors look for that specific case which plays a role in finally detecting what the possible motion could be.

\begin{figure}[!t]
\centering
\includegraphics[width=\columnwidth]{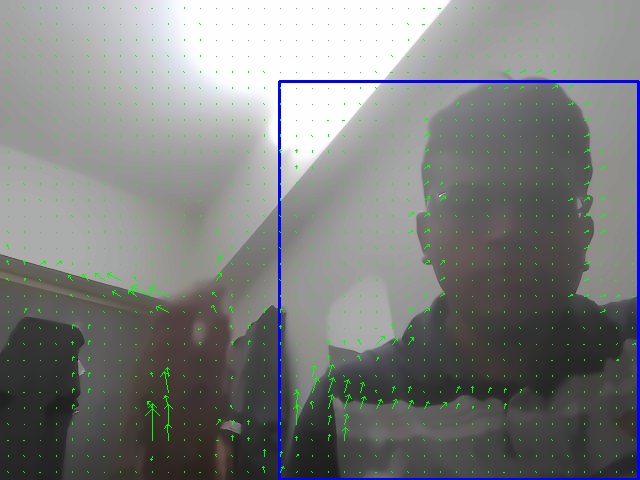}
\caption{For the Case of Finger snapping the green vector flows tell how the objects are moving}
\label{fig:Visual Flow in a test case}
\end{figure}

\subsection{Audio Classification Performance}

We have used real-time UI to show the waveforms change in the case of an Audio change and in the Spectrogram the disturbance is also noted

\begin{figure}[!t]
\centering
\includegraphics[width=\columnwidth]{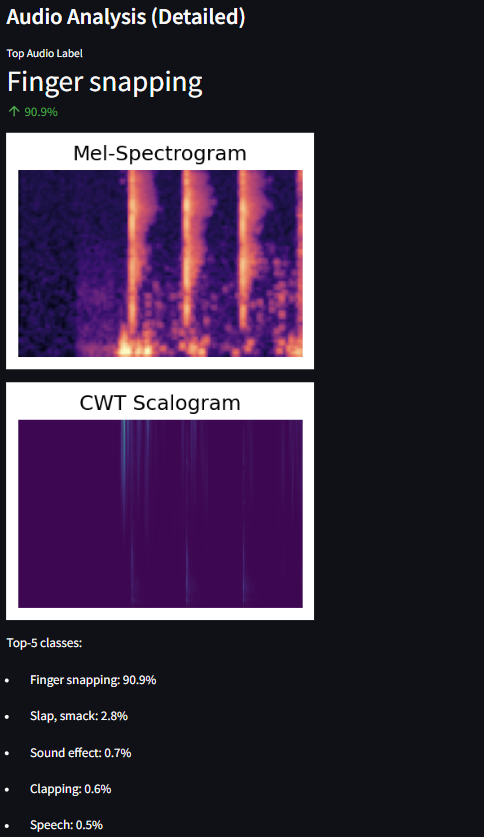}
\caption{Audio classification performance Multi-model ensemble (AST+Wav2Vec2+HuBERT) showing improved accuracy and robustness.}
\label{fig:audio_ensemble}
\end{figure}

Figure~\ref{fig:audio_ensemble} illustrates the performance  achieved through multi-model audio ensemble.

\subsection{Fusion Architecture Evaluation}

Finally the outputs are predicted based on the audio + image inputs to give a better idea of the anomaly happened

\subsection{Anomaly Detection Performance}

It gives the top five possible cases of motion happening as per audio and then uses the visual information to finalize the final cause of the anomaly 

\subsection{Computational Performance}

Mentioned in the Table in detail time taken for each of the processing 

Table~\ref{tab:performance} summarizes the computational performance of system components.

\begin{table}[!t]
\renewcommand{\arraystretch}{1.3}
\caption{Computational Performance Comparison (CPU/GPU)}
\label{tab:performance}
\centering
\begin{tabular}{|c|c|c|}
\hline
\textbf{Component} & \textbf{Basic System (ms)} & \textbf{Advanced System (ms)} \\
\hline
Frame Denoising & 50-100 & 50-100 \\
YOLO Detection & 100-200 & 100-200 \\
DETR Detection & - & 500-1000 \\
ByteTrack Tracking & 10-20 & 10-20 \\
Optical Flow & 200-500 & 200-500 \\
AST Classification & 200-300 & 200-300 \\
Wav2Vec2 Processing & - & 150-250 \\
HuBERT Processing & - & 150-250 \\
Audio Fusion & - & 10-20 \\
Fusion Inference & 5-10 & 15-30 \\
\hline
\textbf{Total per Frame} & \textbf{565-1130} & \textbf{1235-2850} \\
\hline
\end{tabular}
\end{table}

\subsection{Industrial Monitoring Case Study}

Fire case which is a genuine Anomaly case 

\begin{figure}[!t]
\centering
\includegraphics[width=\columnwidth]{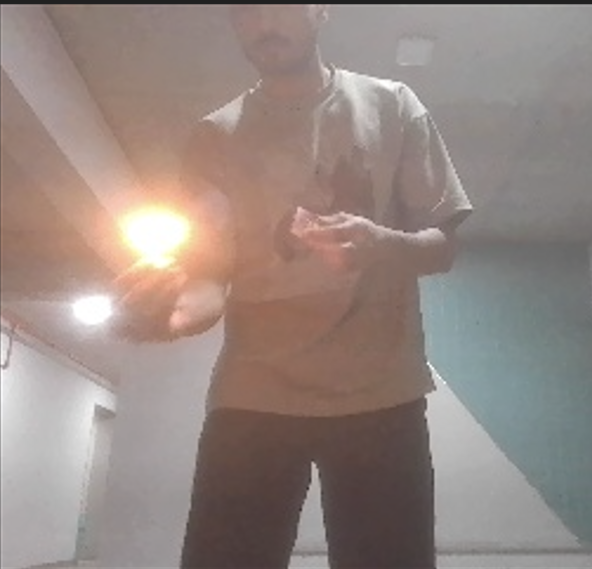}
\caption{Industrial monitoring results we tried to create a fire which was marked as anomaly and its image was also saved in the anomalies section}
\label{fig:industrial_monitoring}
\end{figure}

Figure~\ref{fig:industrial_monitoring} shows example results from industrial monitoring mode.

\begin{figure}[!t]
\centering
\includegraphics[width=\columnwidth]{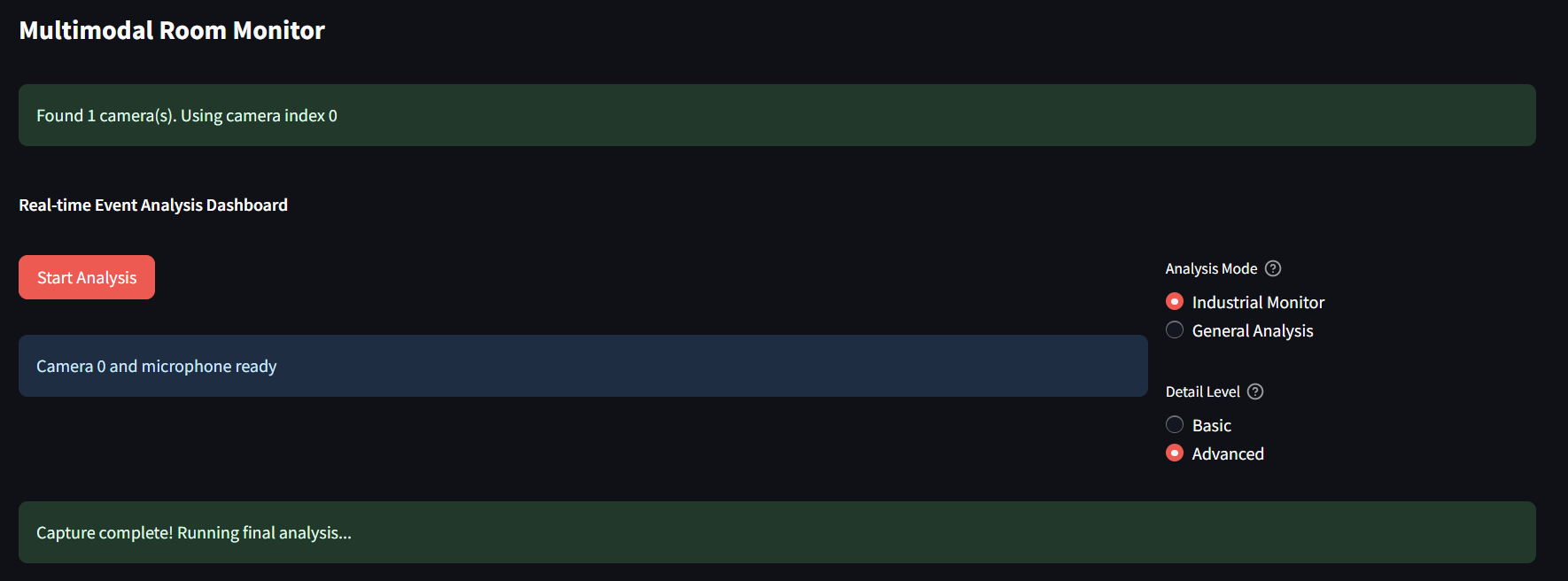}
\caption{UI of the Industrial Analysis}
\end{figure}

\begin{figure}[!t]
\centering
\includegraphics[width=\columnwidth]{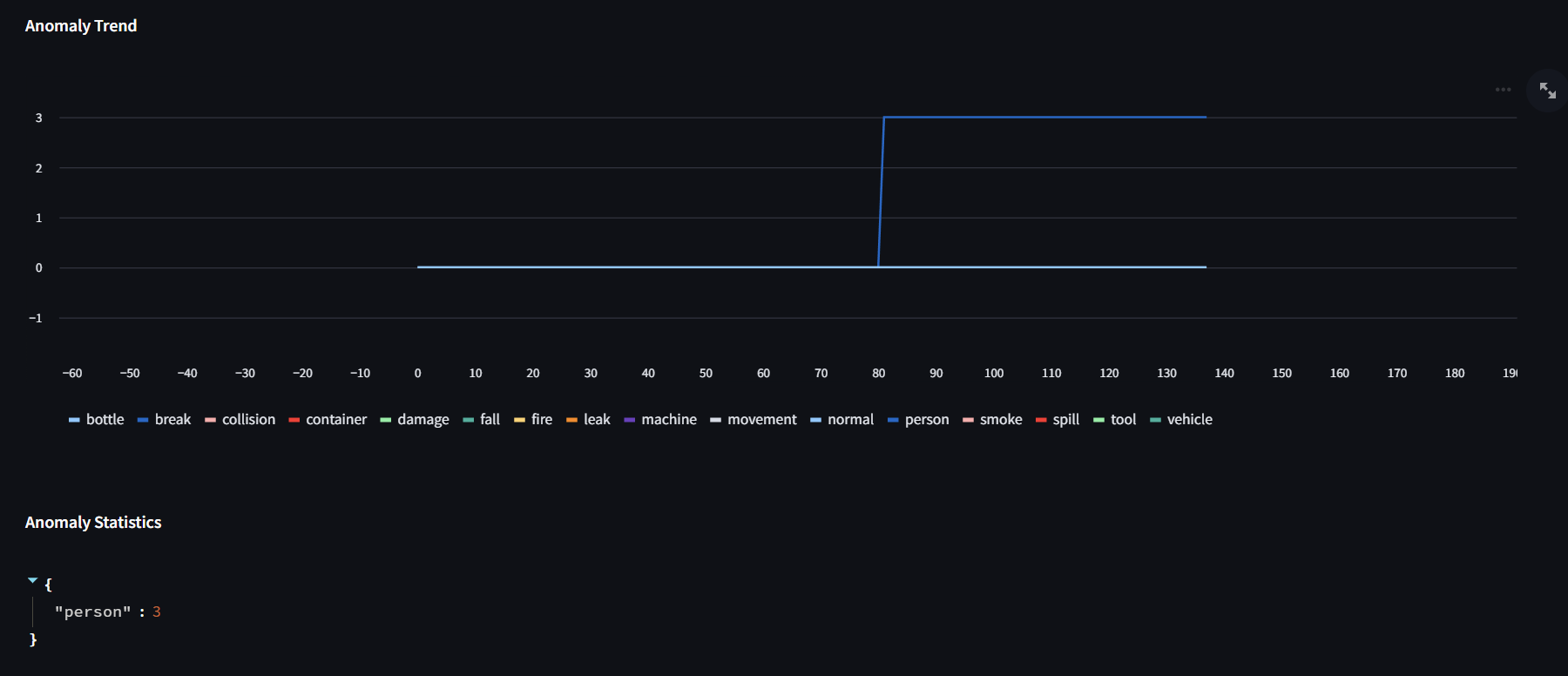}
\caption{Industrial monitoring case graph telling when which anomaly happend along with the time stamp}
\end{figure}

\begin{figure}[!t]
\centering
\includegraphics[width=\columnwidth]{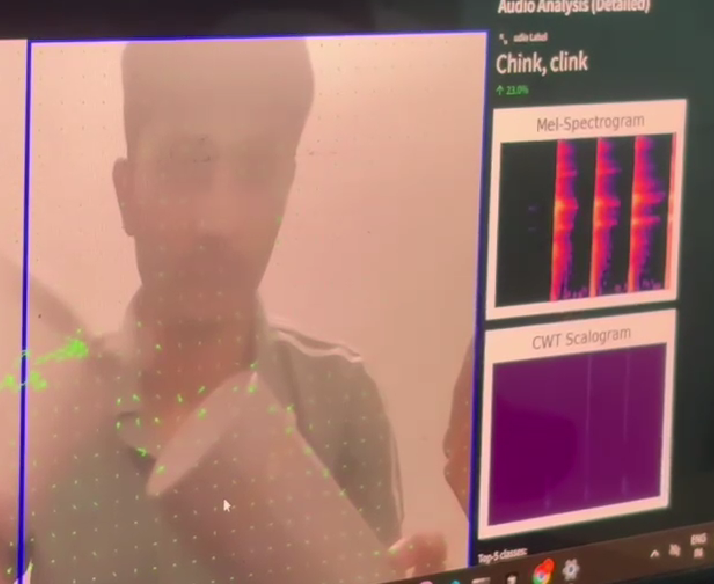}
\caption{Metal Clinking Action and Sound, Common in Industries, Audio Analysis}
\end{figure}

\begin{figure}[!t]
\centering
\includegraphics[width=\columnwidth]{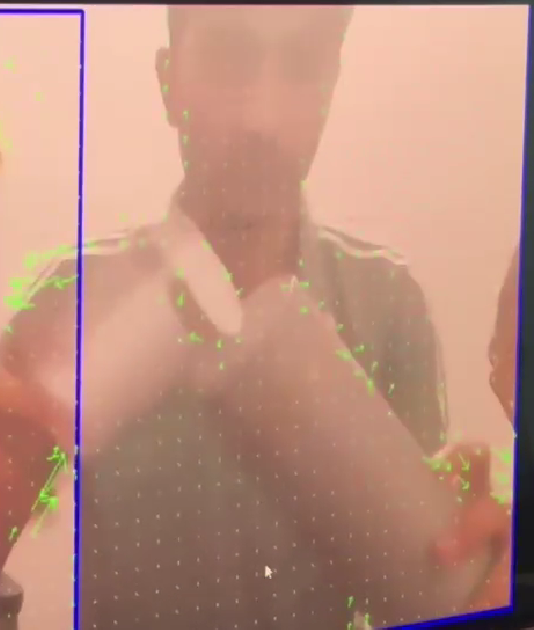}
\caption{Metal Clinking Action and Sound, Common in Industries, Video Analysis, marking the spray motion vectors}
\end{figure}

\begin{figure}[!t]
\centering
\includegraphics[width=\columnwidth]{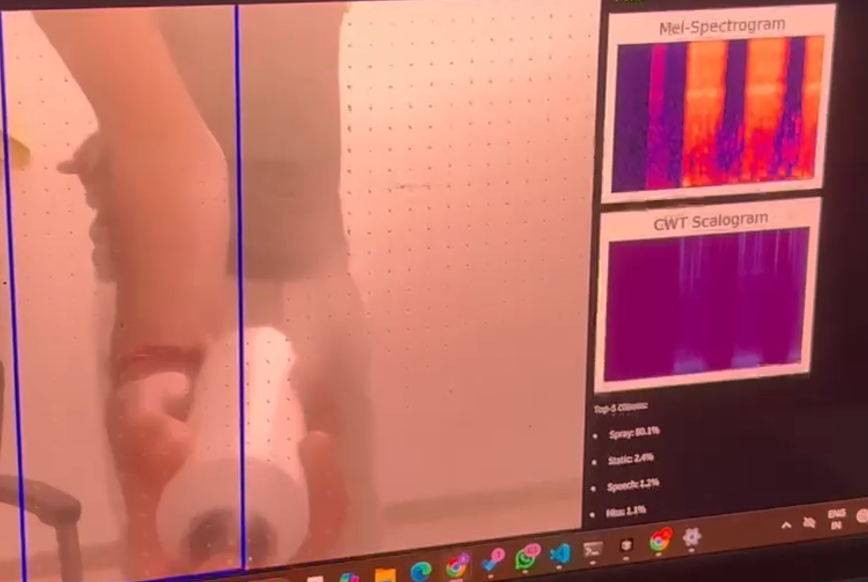}
\caption{Spraying Action and Sound, Common in Industries, Audio Analysis}
\end{figure}

\begin{figure}[!t]
\centering
\includegraphics[width=\columnwidth]{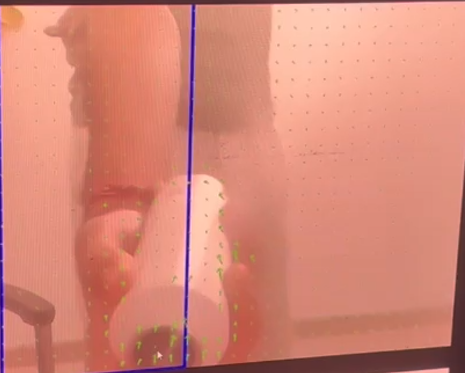}
\caption{Spraying Action and Sound, Common in Industries, Video Analysis marking the spray motion vectors}
\end{figure}

\section{Discussion}

The evolution from basic to advanced system demonstrates significant improvements across multiple dimensions. The multi-model audio ensemble provides more robust audio understanding, with each model contributing specialized knowledge. The hybrid object detection system balances speed and accuracy effectively, addressing practical deployment requirements. Bidirectional cross-modal attention enables richer cross-modal understanding compared to simple fusion methods.

The advanced system maintains real-time performance despite increased complexity, through careful architectural design and optimization. The multi-method anomaly detection approach provides comprehensive coverage, with different methods catching different types of anomalies. Industrial monitoring capabilities extend the system's applicability to safety-critical scenarios.

However, the advanced system requires more computational resources and model storage. The tradeoff between accuracy and efficiency must be carefully considered for specific applications. Future work could explore model compression, knowledge distillation, and efficient attention mechanisms to reduce computational requirements while maintaining accuracy.

\section{Conclusion}

This paper presented the comprehensive evolution of a multimodal room monitoring system from a basic proof-of-concept to an advanced industrial-grade solution. The initial system demonstrated the feasibility of real-time multimodal fusion using efficient models, while the advanced system significantly improved accuracy, robustness, and applicability through multi-model ensembles, hybrid detection, bidirectional attention, and comprehensive anomaly detection.

The key contributions include: (1) detailed architectural evolution demonstrating practical system development, (2) multi-model audio ensemble for robust audio understanding, (3) hybrid object detection balancing speed and accuracy, (4) bidirectional cross-modal attention for enhanced fusion, (5) multi-method anomaly detection suitable for industrial applications, and (6) comprehensive evaluation demonstrating real-world applicability.

Future work will focus on: (1) model compression and efficiency optimization, (2) extension to additional modalities such as depth sensing and thermal imaging, (3) end-to-end trainable architectures, (4) edge device deployment optimization, (5) federated learning for privacy-preserving deployment, and (6) integration with IoT ecosystems for smart environment applications.

\end{document}